\documentclass[aps,prd,onecolumn,eqsecnum,amsmath,nofootinbib,preprintnumbers]{revtex4}%

\usepackage{color,graphicx}
\usepackage{amsfonts,amssymb,theorem,mathrsfs,times}
\usepackage{bm}
{\theorembodyfont{\upshape}
}
{\theorembodyfont{\upshape}
}
{\theorembodyfont{\upshape}
}
{\theorembodyfont{\upshape}
}
{\theorembodyfont{\upshape}
}
{\theorembodyfont{\upshape}
}

\newcommand{\dalm}{\kern1pt\vbox{\hrule height 0.9pt\hbox{\vrule width
0.9pt\hskip 2.5pt\vbox{\vskip 5.5pt}\hskip 3pt\vrule width
0.3pt}\hrule height 0.3pt}\kern1pt}

\begin{document}
\preprint{\hfill {\small {ICTS-USTC-15-07}}}
\title{$P$-$V$ criticality in the extended phase space of black holes in massive gravity}

%

\author{Jianfei Xu$^a$\footnote{e-mail address: jfxu06@mail.ustc.edu.cn}, Li-Ming Cao$^{a,b}$\footnote{e-mail
address: caolm@ustc.edu.cn}, Ya-Peng Hu$^{b,c}$\footnote{e-mail address: huyp@nuaa.edu.cn}}


\affiliation{$^a$
Interdisciplinary Center for Theoretical Study\\
University of Science and Technology of China, Hefei, Anhui 230026,
China}

\affiliation{$^b$ State Key Laboratory of Theoretical Physics,
Institute of Theoretical Physics, Chinese Academy of Sciences, P.O.
Box 2735, Beijing 100190, China}

\affiliation{$^c$ College of Science, Nanjing University of
Aeronautics and Astronautics, Nanjing 210016, China}

\date{\today}

\begin{abstract}
We study the $P$-$V$ criticality and phase transition in the extended
phase space of charged anti-de Sitter black holes in canonical ensemble of
ghost-free massive gravity, where the cosmological constant is
viewed as a dynamical pressure of the black hole system. We give the generalized thermodynamic first law and the Smarr relation with
massive gravity correction. We find that not only when the horizon
topology is spherical but also in the Ricci flat or hyperbolic case,
there appear the $P$-$V$ criticality and phase transition up to the
combination $k+c_0^2c_2m^2$ in the four-dimensional case, where $k$ characterizes the horizon curvature and $c_2m^2$ is the coefficient of the second term of massive potential associated with the graviton mass. The positivity of such combination indicate the van der Waals-like phase transition.  When the spacetime dimension is larger then four, the Maxwell charge there seems unnecessary for the appearance of critical behavior, but a infinite repulsion effect needed, which can also be realized through negative valued $c_3m^2$ or $c_4m^2$, which is third or fourth term of massive potential. When $c_3m^2$ is positive, a Hawking-Page like black hole to vacuum phase transition is shown in the five-dimensional chargeless case. For the van der Waals like phase transition in four and five spacetime dimensions, we calculate the critical exponents near critical point and find they are the same as those in the van der Waals liquid-gas system. 
\end{abstract}


\maketitle


\section{Introduction}
Over the past decades, the thermodynamics of black holes in
asymptotically anti-de Sitter (AdS) space has been widely explored
in contrast with its counterpart in asymptotically flat or de Sitter
(dS) space since the AdS black holes are thermodynamically
stable~\cite{Hawking:1982dh}. Another interesting reason for the focus on AdS
black holes is the advent of AdS/CFT 
correspondence~\cite{Maldacena:1997re, Gubser:1998bc,
Witten:1998qj}, which relates a gravity theory in the bulk of an
asymptotical AdS space to a conformal field theory without gravity
on the boundary of AdS space. According to this holographic
dictionary, a bulk AdS black hole corresponds to a boundary finite
temperature conformal field theory. At this stage, we can understand
a strongly coupled field theory by studying a weakly coupled
classical gravity theory. For example, AdS black holes can undergo
a so-called Hawking-Page phase transition to a thermal AdS space
under a certain Hawking temperature~\cite{Hawking:1982dh}. This
phenomenon in gravity theory can be explained as the
confinement or deconfinement phase transition in large-$N$ gauge
theory~\cite{Witten:1998zw}. More interesting phenomena show up when
we consider the charged AdS black holes~\cite{Chamblin:1999tk,
Chamblin:1999hg}. There exists a first-order phase transition between
large black holes and small black holes when the charge is below a
critical value. Such phase transitions and critical behaviors of
classically charged black holes in AdS space are very similar to a
van der Waals liquid-gas phase transition.

Although the thermodynamics of charged black holes in AdS space is
similar to the van der Waals system, it should be noted that the
corresponding critical behaviors appear in the $Q-\Phi$ diagram, where
$Q$ is the charge of the black hole and $\Phi$ is the chemical potential
conjugate to the charge~\cite{Shen:2005nu}. No pressure $P$ or
volume $V$ are defined in such a black hole system, while we use the $P$-$V$
diagram to characterize the van der Waals liquid-gas phase
transition. Thus, the analogy is problematic since the charge $Q$ is an
extensive quantity and $\Phi$ is an intensive one in the black hole
thermodynamics, while $P$ is an intensive quantity and $V$ is an
extensive one in the van der Waals system. The way to solve this
problem is by including the variation of the cosmological constant
$\Lambda$ in the first law of black hole
thermodynamics~\cite{Caldarelli:1999xj, Kastor:2009wy, Dolan:2010ha,
Dolan:2011xt, Dolan:2011jm, Cvetic:2010jb, Lu:2012xu}. Since the
dimension of the cosmological constant over the Newtonian constant
$\Lambda/G_N$ is equal to the dimension of pressure, it is natural
to identify the cosmological constant as the thermodynamical
pressure of the system ($G_N=\hbar=c=k=1$),
\begin{equation}
P=-\frac{1}{8\pi}\Lambda=\frac{n(n+1)}{16\pi l^2},
\end{equation}
in $n+2$-dimensional spacetime. There are some physical reasons for
this identification~\cite{Kubiznak:2012wp}. First, one can imagine
that there exist ``more fundamental'' theories, where the physical
constant, such as the Yukawa coupling, gauge coupling constants,
Newtonian constant, or the cosmological constant arise as vacuum
expectation values and, hance, can vary. Second, the Smarr relation
obtained by the scaling method becomes inconstant with the first law of
black hole thermodynamics unless the variation of $\Lambda$ is
included~\cite{Kastor:2009wy}. In addition, once one views the
cosmological constant as the thermodynamical pressure, the black
hole mass $M$ should be understood as enthalpy rather than the internal
energy of the system~\cite{Kastor:2009wy}. Keeping this in mind, we
can use the standard thermodynamic identity to get ``thermodynamic
volume'' which is conjugate to the pressure of black
holes~\cite{Parikh:2005qs, Ballik:2013uia, Kubiznak:2012wp}. In this
way, people can investigate the $P$-$V$ critical behaviors of AdS
black holes and find exactly the same behaviors as in the van der Waals
liquid-gas system. The $P$-$V$ criticality study of AdS black holes
has now been pushed even further to other gravities such as higher-derivative gravities~\cite{Wei:2012ui, Xu:2014tja, Xu:2014kwa,
Cai:2013qga}.

In the above discussion, the gravity theories are either Einstein's
general relativity (GR) which has been widely accepted as a correct
theory of gravity at low energies or the other possible covariant
gravity theories. The most important principle of GR is that it is a
theory of a nontrivially interacting massless helicity-2
particle--massless graviton. However, GR is not UV complete. It must
be a effective field theory valid at energy up to a cutoff at
a certain energy scale beyond which high-energy effects will take place
and the gravity theory should be
modified~\cite{Hinterbichler:2011tt}. Massive gravity, where
the graviton is endowed with mass, is one of the most straightforward
modifications of GR. The construction of the linear theory of massive
gravity was first given by Fierz and Pauli in
1939~\cite{Fierz:1939}. While at the nonlinear level, the
traditional constructions of massive gravity are plagued by the
Boulware-Deser (BD) ghost instability~\cite{Boulware:1972zf,
Boulware:1973my}, up to now, much progress has been made in
overcoming the ghost instability. The ghost-free massive gravity was
also proposed recently~\cite{Hassan:2011hr, Hassan:2011tf}. More
recently, a nontrivial black hole solution was found in ghost-free
massive gravity with a negative cosmological
constant~\cite{Vegh:2013sk}. These black holes' corresponding
thermodynamical properties and phase structure have been studied
by~\cite{Cai:2014znn}. The aim of this paper is to analyze the
extended phase structure and investigate the $P$-$V$ critical behavior
of charged AdS black holes in canonical ensemble in that ghost-free
massive gravity.

This paper is organized as follows. In Sec.II, we present the
thermodynamics of massive gravity black holes. We extend the phase
space by viewing the cosmological constant as the thermodynamical
pressure of the black hole system. In Sec.III, we study the $P$-$V$
criticality in four-dimensional case. By properly defined
thermodynamical quantities, a van der Waals-like phase transition is
found and the corresponding critical exponents are calculated. In
Sec.IV, We explore the five-dimensional case, where we find that the
coefficients of massive potential can also play the role of a Maxwell
charge. A Hawking-Page-like phase transition is presented. In Sec.V, we try to understand the physical origin of such
van der Waals-like phase transitions in a massive black hole system.
The last section is devoted to some conclusions and discussion.

\section{Thermodynamics of black holes in massive gravity}
We are considering the following action for an ($n+2$)-dimensional
massive gravity~\cite{Cai:2014znn},
\begin{equation}
S=\frac{1}{16\pi}\int
d^{n+2}x\sqrt{-g}\Big[R+\frac{n(n+1)}{l^2}-\frac{1}{4}F^2+m^2\sum_{i=1}^4c_i\mathcal{U}_i(g,f)\Big],
\end{equation}
where the last four terms are the massive potential associate with
graviton mass, $c_i$ are the constants, $f$ is a fixed symmetric tensor
called the reference metric, and $\mathcal{U}_i$ are symmetric
polynomials of the eigenvalue of the ($n+2$)$\times$($n+2$) matrix
$\mathcal{K}^{\mu}_{~\nu}\equiv\sqrt{g^{\mu\alpha}f_{\alpha\nu}}$:
\begin{eqnarray}
&~&\mathcal{U}_1=[\mathcal{K}],\nonumber\\
&~&\mathcal{U}_2=[\mathcal{K}]^2-[\mathcal{K}^2],\nonumber\\
&~&\mathcal{U}_3=[\mathcal{K}]^3-3[\mathcal{K}][\mathcal{K}^2]+2[\mathcal{K}^3],\nonumber\\
&~&\mathcal{U}_4=[\mathcal{K}]^4-6[\mathcal{K}^2][\mathcal{K}]^2+8[\mathcal{K}^3][\mathcal{K}]+3[\mathcal{K}^2]^2-6[\mathcal{K}^4].
\end{eqnarray}
The square root in $\mathcal{K}$ is understood as the matrix square
root, i.e.,
$(\sqrt{A})^{\mu}_{~\nu}(\sqrt{A})^{\nu}_{~\lambda}=A^{\mu}_{~\nu}$,
and the rectangular brackets denote traces--
$[\mathcal{K}]=\mathcal{K}^{\mu}_{~\mu}$.

The action admits a static black hole solution with the spacetime metric
and reference metric as
\begin{equation}
ds^2=-f(r)dt^2+f^{-1}(r)dr^2+r^2h_{ij}dx^idx^j,
\end{equation}
\begin{equation}\label{refmetric}
f_{\mu\nu}=\mathrm{diag}(0,0,c_0^2h_{ij}),
\end{equation}
where $c_0$ is a positive constant, and $h_{ij}dx^idx^j$ is the line
element for an Einstein space with constant curvature $n(n-1)k$.
Without loss of generality, one may take $k=1$, $0$, or $-1$,
corresponding to a spherical, Ricci flat, or hyperbolic topology of
the black hole horizon, respectively. According to the reference
metric (\ref{refmetric}), we have
\begin{eqnarray}\label{utermeincoor}
&~&\mathcal{U}_1=nc_0/r,\nonumber\\
&~&\mathcal{U}_2=n(n-1)c_0^2/r^2,\nonumber\\
&~&\mathcal{U}_3=n(n-1)(n-2)c_0^3/r^3,\nonumber\\
&~&\mathcal{U}_4=n(n-1)(n-2)(n-3)c_0^4/r^4.
\end{eqnarray}
The metric function $f(r)$ is given by~\cite{Cai:2014znn}
\begin{eqnarray}\label{f}
f(r)&=&k+\frac{16\pi P}{(n+1)n}r^2-\frac{16\pi
M}{nV_nr^{n-1}}+\frac{(16\pi Q)^2}{2n(n-1)V_n^2r^{2(n-1)}}+\frac{c_0c_1m^2}{n}r+c_0^2c_2m^2\nonumber\\
&~&+\frac{(n-1)c_0^3c_3m^2}{r}+\frac{(n-1)(n-2)c_0^4c_4m^2}{r^2},
\end{eqnarray}
where $V_n$ is the volume of space spanned by coordinates $x^i$, $M$
is the black hole mass, $Q$ is related to the charge of the black
hole, and $P=\frac{n(n+1)}{16\pi l^2}$ is the pressure. The black
hole horizon is determined by $f(r)|_{r=r_h}=0$. Thus, the mass $M$
can be expressed in terms of $r_h$ as
\begin{eqnarray}\label{enthalpy}
M&=&\frac{nV_nr_h^{n-1}}{16\pi}\Big[k+\frac{16\pi
P}{(n+1)n}r_h^2+\frac{(16\pi Q)^2}{2n(n-1)V_n^2r_h^{2(n-1)}}+\frac{c_0c_1m^2}{n}r_h+c_0^2c_2m^2\nonumber\\
&~&+\frac{(n-1)c_0^3c_3m^2}{r_h}+\frac{(n-1)(n-2)c_0^4c_4m^2}{r_h^2}\Big].
\end{eqnarray}
The Hawking temperature of the black hole can be easily obtained by
requiring the absence of conical singularity at the horizon in the
Euclidean sector of the black hole solution, which is given by
\begin{eqnarray}\label{Htemp}
T=\frac{1}{4\pi}f'(r_h)&=&\frac{1}{4\pi r_h}\Big[(n-1)k+\frac{16\pi
P}{n}r_h^2-\frac{(16\pi Q)^2}{2nV_n^2r_h^{2(n-1)}}+c_0c_1m^2r_h+(n-1)c_0^2c_2m^2\nonumber\\
&~&+\frac{(n-1)(n-2)c_0^3c_3m^2}{r_h}+\frac{(n-1)(n-2)(n-3)c_0^4c_4m^2}{r_h^2}\Big].
\end{eqnarray}
We are now going to discuss the thermodynamics of black holes in
massive gravity in extended phase space by introducing the pressure
$P=\frac{n(n+1)}{16\pi l^2}$; the black hole mass $M$ should be
viewed as the enthalpy $H\equiv M$ rather than the internal energy of the
gravitational system~\cite{Kastor:2009wy}. The other thermodynamic
quantities can be obtained through thermodynamic identities. The
entropy $S$, thermodynamic volume $V$, and electric potential $\Phi$
are given by
\begin{equation}
S=\int_0^{r_h}T^{-1}\left(\frac{\partial H}{\partial
r}\right)_{Q,P}dr=\frac{V_n}{4}r_h^n,
\end{equation}
\begin{equation}\label{thermovolume}
V=\left(\frac{\partial H}{\partial
P}\right)_{S,Q}=\frac{V_n}{n+1}r_h^{n+1},
\end{equation}
\begin{equation}
\Phi=\left(\frac{\partial H}{\partial
Q}\right)_{S,P}=\frac{16\pi}{(n-1)V_nr_h^{n-1}}Q.
\end{equation}
It is easy to check that those thermodynamic quantities obey the
following differential equation,
\begin{equation}\label{firstlaw}
\mathrm{d}H=T\mathrm{d}S+V\mathrm{d}P+\Phi\mathrm{d}Q+\frac{V_nc_0m^2r_h^n}{16\pi}\mathrm{d}c_1+\frac{nV_nc_0^2m^2r_h^{n-1}}{16\pi}\mathrm{d}c_2+\frac{n(n-1)V_nc_0^3m^2r_h^{n-2}}{16\pi}\mathrm{d}c_3+\frac{n(n-1)(n-2)V_nc_0^4m^2r_h^{n-3}}{16\pi}\mathrm{d}c_4,
\end{equation}
where we have viewed the coupling constants $c_i$ as variables. Invoking the scaling method, we can get the Smarr relation for the
black hole with $c_i$ terms as
\begin{equation}
(n-1)H=nTS-2PV+(n-1)\Phi
Q-\frac{V_nc_0c_1m^2}{16\pi}r_h^n+\frac{n(n-1)V_nc_0^3c_3m^2}{16\pi}r_h^{n-2}+\frac{n(n-1)(n-2)V_nc_0^4c_4m^2}{8\pi}r_h^{n-3}.
\end{equation}

\section{$P$-$V$ criticality of four-dimensional black holes}
In the four-dimensional spacetime case, we can simply set $c_3=c_4=0$
since we have $\mathcal{U}_3=\mathcal{U}_4=0$ according to
Eq.(\ref{utermeincoor}) when $n=2$. In this case, the enthalpy
(\ref{enthalpy}) is reduced to
\begin{equation}
H=\frac{V_2r_h}{8\pi}\Big[k+\frac{8\pi P}{3}r_h^2+\frac{(8\pi
Q)^2}{V_2^2r_h^2}+\frac{c_0c_1m^2}{2}r_h+c_0^2c_2m^2\Big].
\end{equation}
And the equation of state of the black holes can be obtained from the
Hawking temperature (\ref{Htemp}) in this four-dimensional case as
\begin{equation}\label{P}
P=\Big(\frac{T}{2}-\frac{c_0c_1m^2}{8\pi}\Big)\frac{1}{r_h}-\Big(\frac{k}{8\pi}+\frac{c_0^2c_2m^2}{8\pi}\Big)\frac{1}{r_h^2}+\frac{8\pi
Q^2}{V_2^2}\frac{1}{r_h^4}.
\end{equation}
We are now going to study the phase structure of black holes in the
canonical ensemble with fixed charge in terms of $P$-$V$ diagram. Note
that the thermodynamic volume $V$ (\ref{thermovolume}) is a
monotonic function of the horizon radius $r_h$, so we can use $r_h$
to specify the critical behavior instead of $V$. The critical point
is determined as the inflection point in the $P$-$V$ diagram, i.e.,
\begin{equation}\label{cricondition}
\frac{\partial P}{\partial
r_h}\Big|_{r_h=r_{hc},T=T_c}=\frac{\partial^2P}{\partial
r_h^{~2}}\Big|_{r_h=r_{hc},T=T_c}=0.
\end{equation}
For further convenience, we would like to denote the coefficients in
the equation of state (\ref{P}) as
\begin{eqnarray}\label{w124}
&~&w_1=\frac{T}{2}-\frac{c_0c_1m^2}{8\pi}\nonumber\\
&~&w_2=-\Big(\frac{k}{8\pi}+\frac{c_0^2c_2m^2}{8\pi}\Big)\nonumber\\
&~&w_4=\frac{8\pi Q^2}{V_2^2}.
\end{eqnarray}
From now on, we shall denote $w_1$ as half of the effective temperature
or shifted temperature, for short. Obviously, such a temperature can be negative according to the value of $c_1m^2$. Thus, the critical point determined by Eq.(\ref{cricondition}) with critical quantities can be calculated as
\begin{equation}
r_{hc}=\sqrt{-\frac{6w_4}{w_2}},
\end{equation}
\begin{equation}
w_{1c}=-\frac{4}{3}w_2\sqrt{-\frac{w_2}{6w_4}},
\end{equation}
\begin{equation}
P_c=\frac{w_2^2}{12w_4}.
\end{equation}
Note that the critical behavior occurs only when $w_2<0$. We can
easily find a universal relation among critical pressure $P_c$,
shifted temperature $w_{1c}$, and horizon radius $r_{hc}$,
\begin{equation}
\frac{P_cr_{hc}}{w_{1c}}=\frac{3}{8}.
\end{equation}
This is called the critical coefficient, which keeps the same value
as the van der Waals liquid-gas system. However, if we use real the
critical temperature $T_c$ instead, we cannot get a constant
critical coefficient.

The $P$-$V(r_h)$ diagrams have been drawn in the Fig. 1 with
different shifted temperature $w_1$. The left plot shows exactly the
same behavior as the van der Waals system and has the critical point,
while the right plot shows the monotone phase structure according to the
different sign of $w_2$. Thus, changing the sign of $w_2$ can
dramatically change the phase diagram. The van der Waals-like phase
diagram indicates that there must exist a phase transition when
$w_2<0$. To clearly specify the phase transition, we shall introduce
the Gibbs free energy as a Legendre transformation of enthalpy as
\begin{equation}\label{G}
G=H-TS.
\end{equation}
By using Eq.(\ref{firstlaw}), we can immediately get the exterior
derivative of Gibbs free energy as
$\mathrm{d}G=-S\mathrm{d}T+V\mathrm{d}P+\Phi\mathrm{d}Q$, which tell
us that the Gibbs free energy will take the minimal value in the equilibrium
state when the temperature, pressure, and charge of the system are all
held fixed. In Fig. 2, we have plotted the Gibbs free energy
as a function of the shifted temperature for various pressures. It can
be seen from the left diagram that when $w_2<0$ and $P<P_c$, there exists a ``swallow tail''-type  behavior and indicate a first-order
phase transition as expected. The ``tail'' characterizes the unstable
state, since the Gibbs-free energy always takes the minimal value
for constant temperature, pressure, and electric charge. The
``swallow tail'' disappears when $P>P_c$. The right diagram is
monotonically decreasing, and has no conflict with $P$-$V(r_h)$ statement.
There exists no phase transition when $w_2>0$. We can see from the $G$-$w_1$ diagrams that the shifted temperature can take a negative value for a stable black hole phase. However, if we plot the $P$-$V(r_h)$ diagram with negative $w_1$, we will find that the stable black hole will have a maximal horizon radius above which the pressure will become negative.

\begin{figure}
\includegraphics[width=8cm]{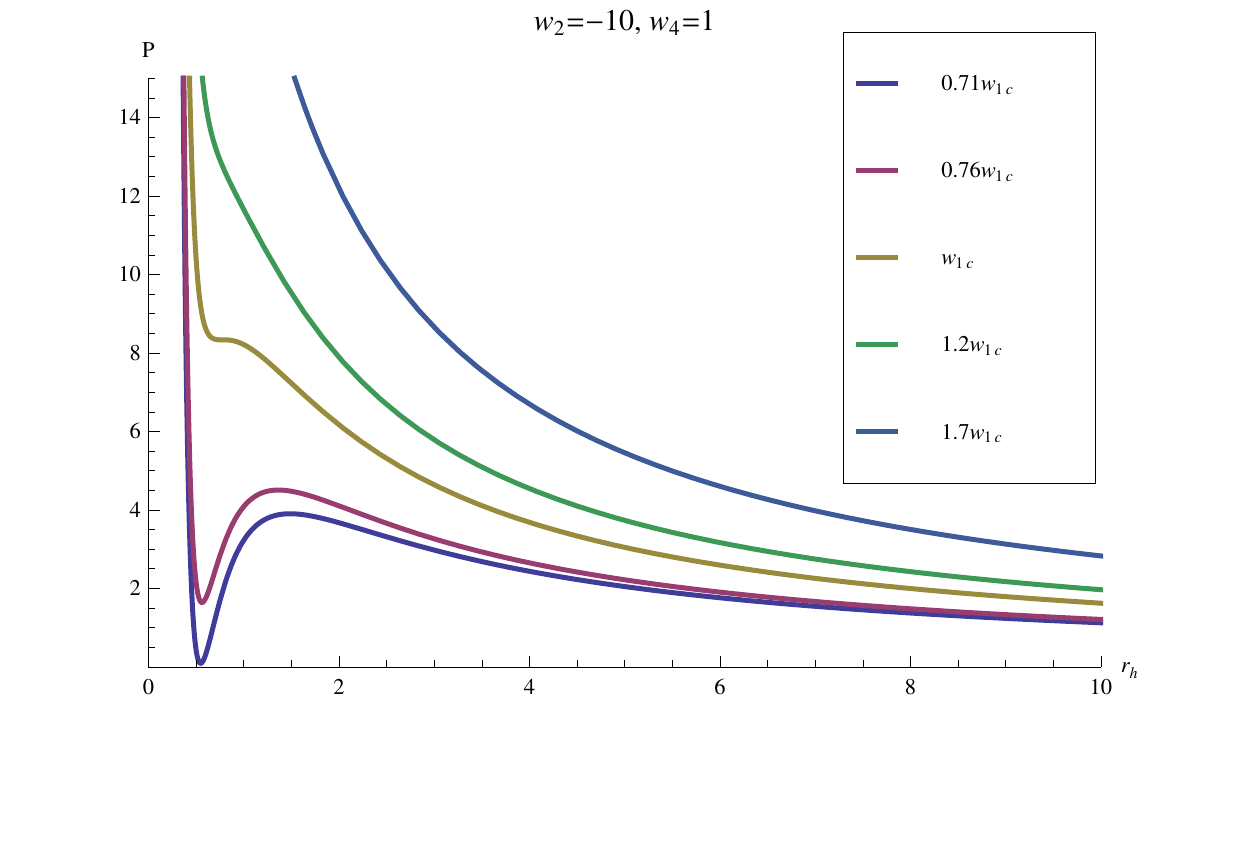}
\includegraphics[width=8cm]{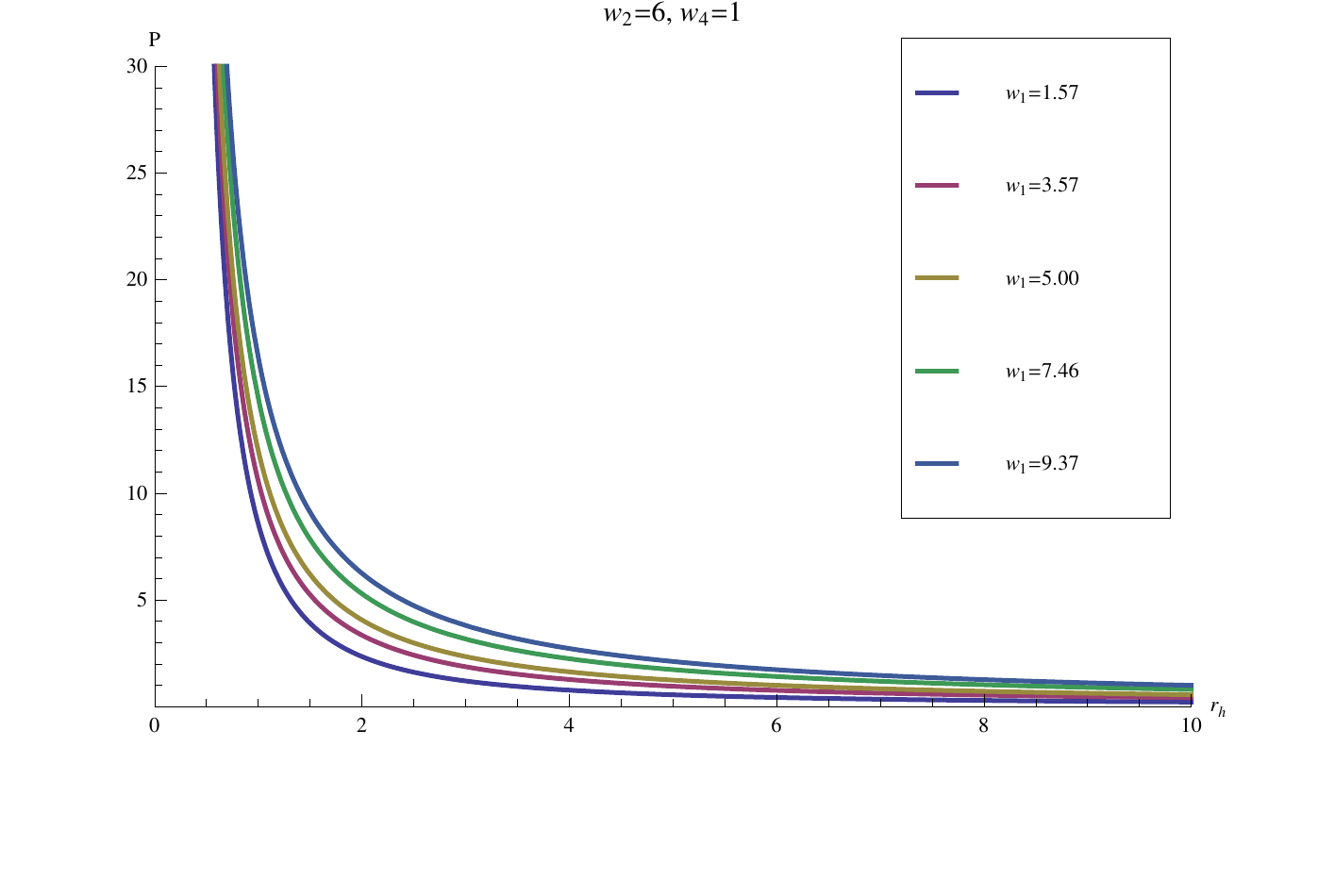}
\caption{The four-dimensional $P$-$V(r_h)$ diagrams for $w_2=-10$ and $w_2=6$, where we have set $w_4=1$ for simplicity. The left diagram shows the van der Waals-like critical behavior, while the right one does not according to the different sign of $w_2$.}
\end{figure}

\begin{figure}
\includegraphics[width=8cm,height=6cm]{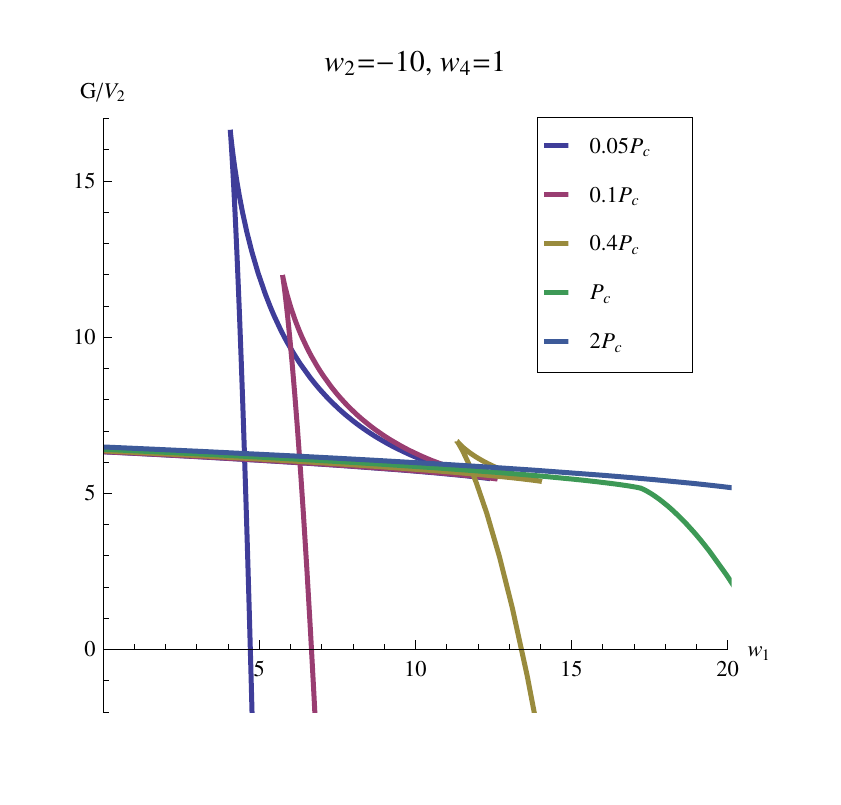}
\includegraphics[width=8cm,height=6cm]{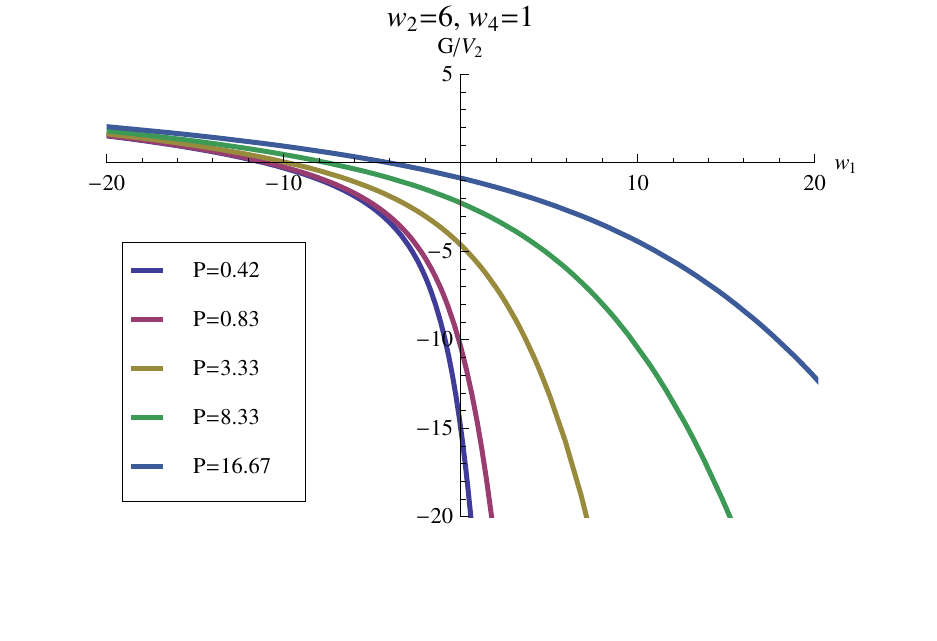}
\caption{The four-dimensional Gibbs free energy as a function of shifted
temperature for different pressures. The left diagram shows the
``swallow tail'' behavior while the right one does not according to
the different sign of $w_2$.}
\end{figure}

\begin{figure}
\includegraphics[width=8cm]{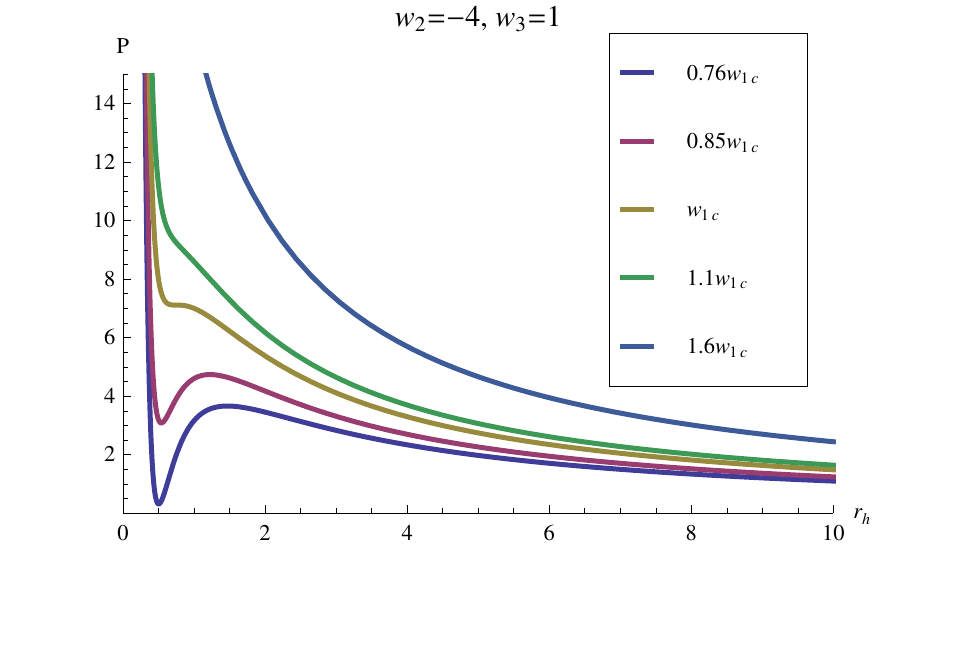}
\includegraphics[width=8cm]{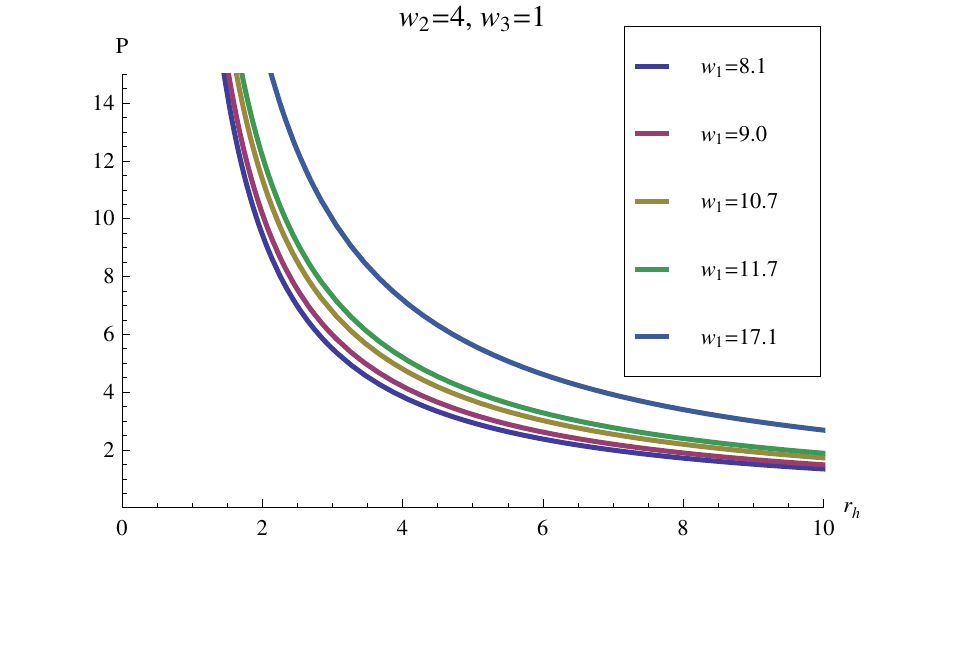}\\
\includegraphics[width=8cm]{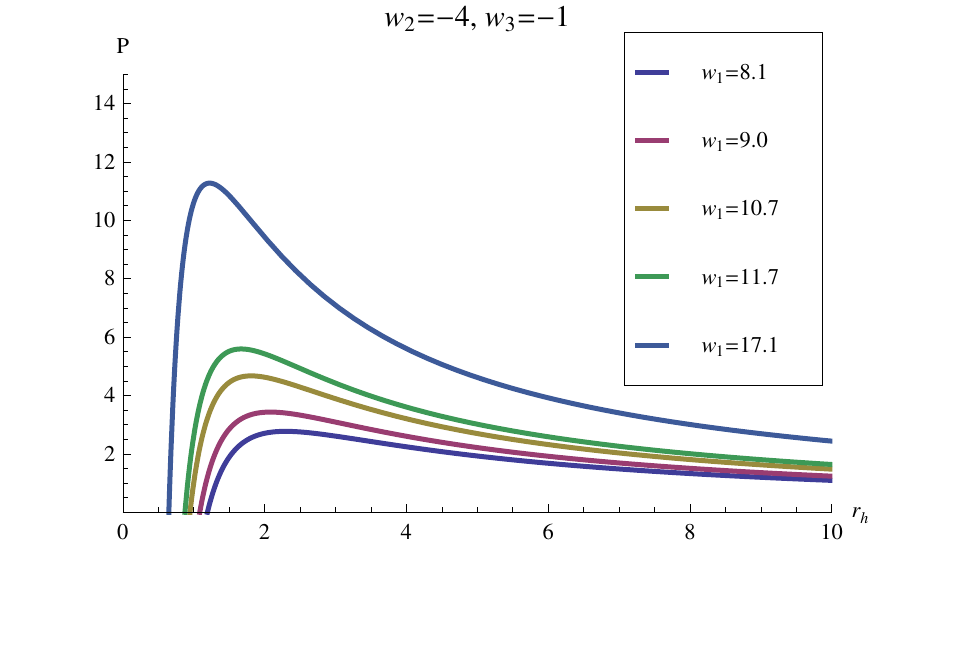}
\includegraphics[width=8cm]{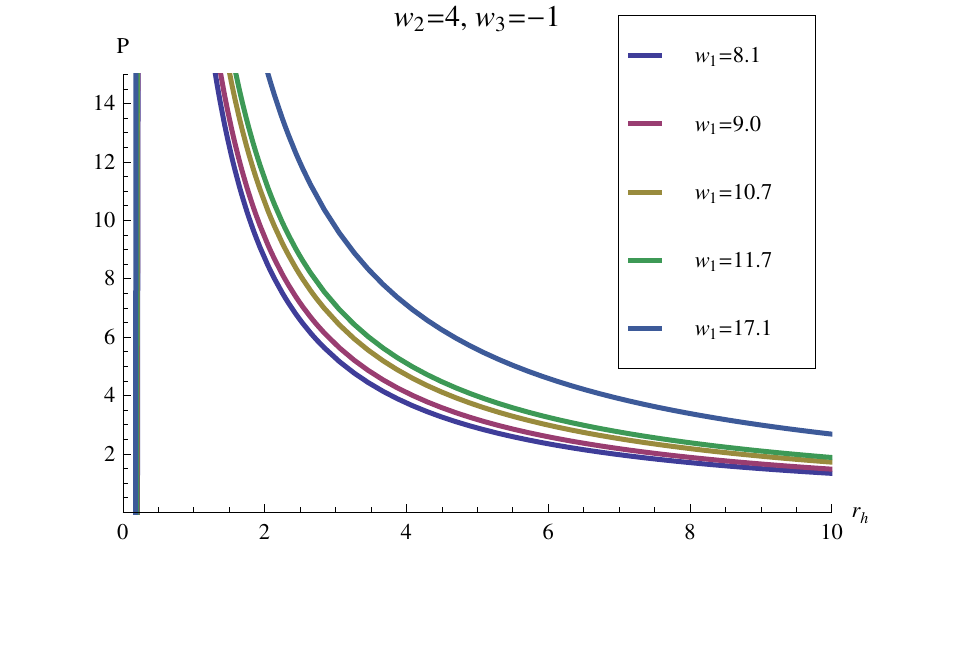}
\caption{The five-dimensional $P$-$V(r_h)$ diagrams with different
parametrization. The first diagram shows the van der Waals-like
critical behavior when $w_2<0$ and $w_3>0$.}
\end{figure}

\begin{figure}
\includegraphics[width=8cm,height=6cm]{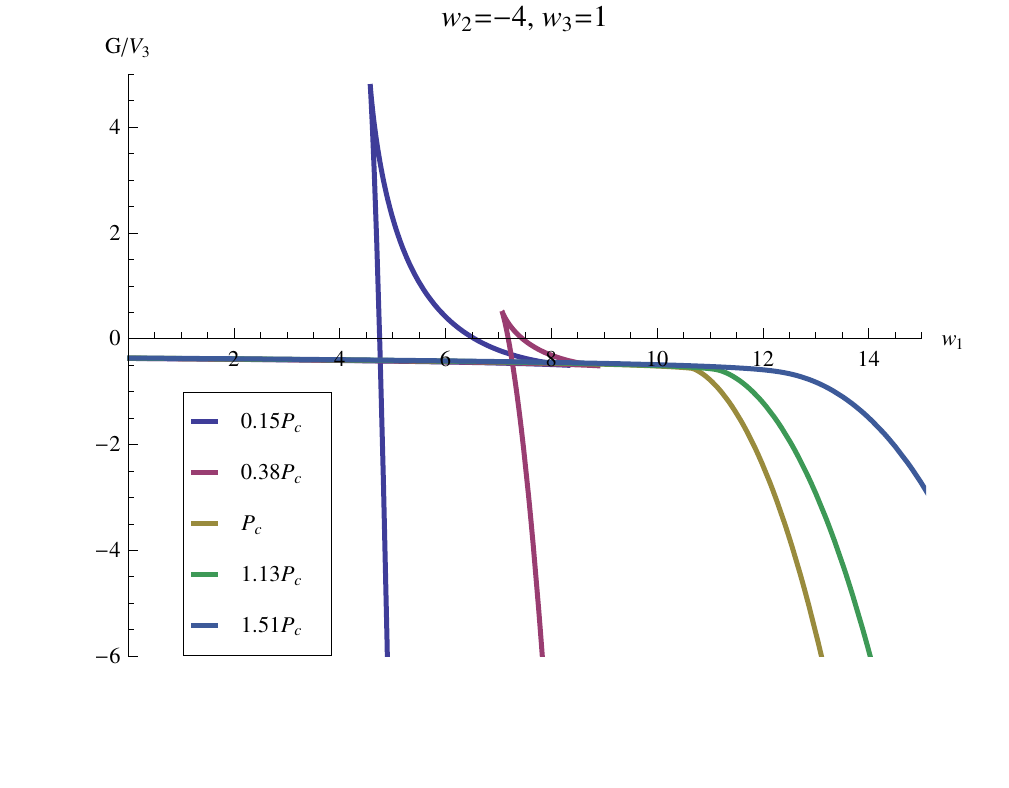}
\includegraphics[width=8cm,height=6cm]{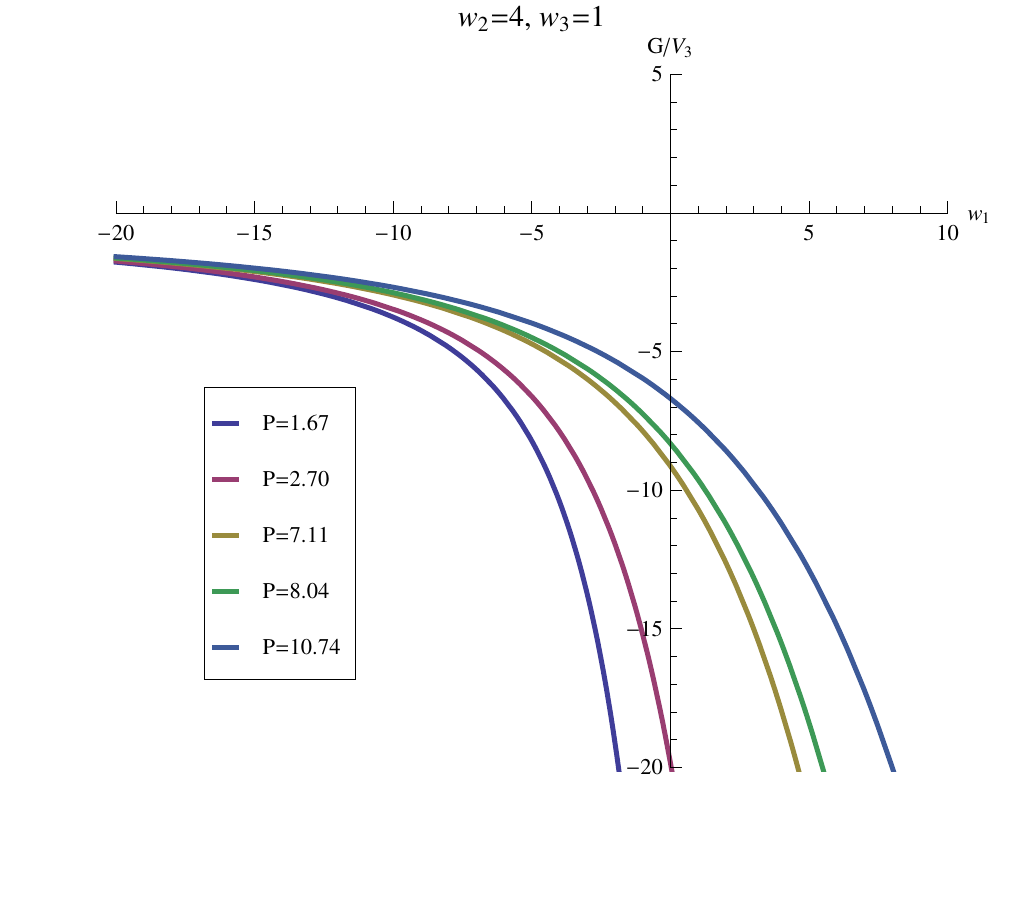}\\
\includegraphics[width=8cm,height=6cm]{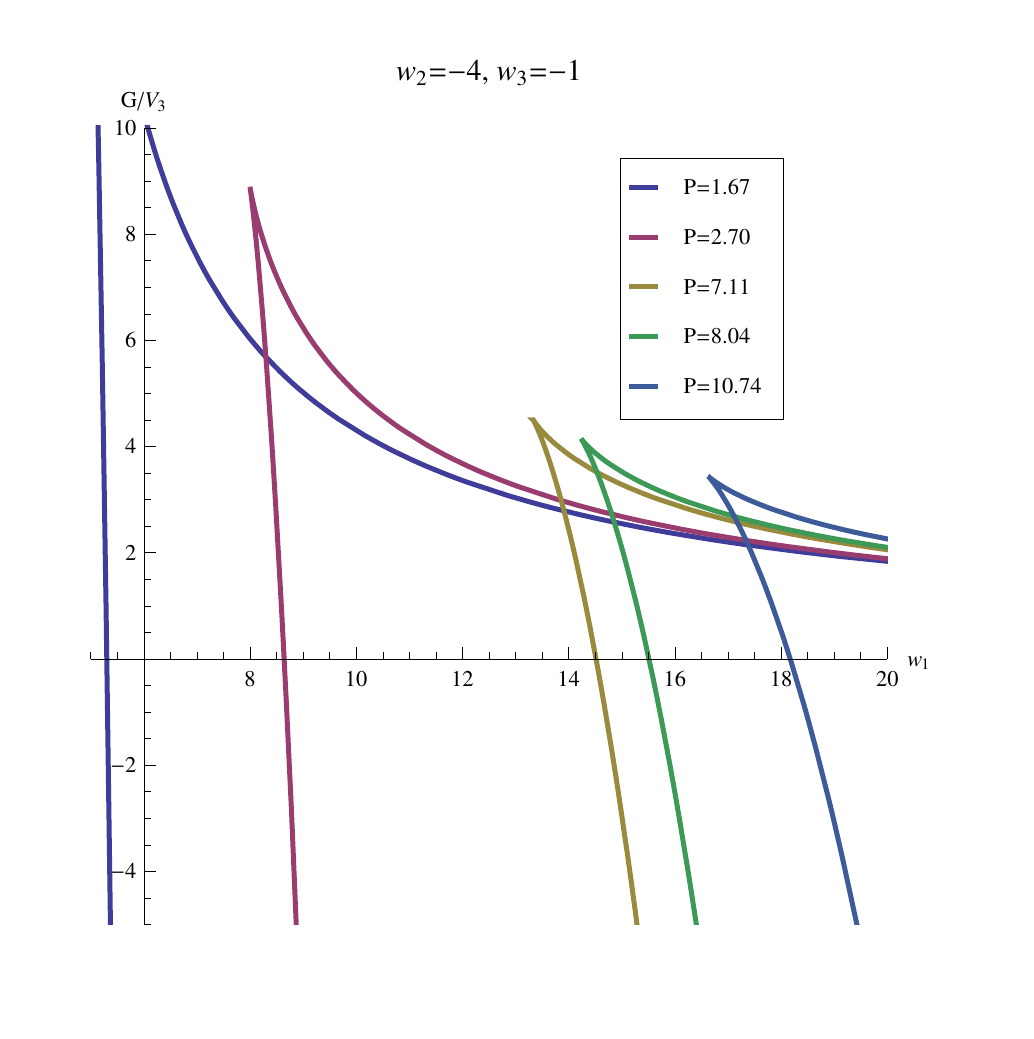}
\includegraphics[width=8cm,height=6cm]{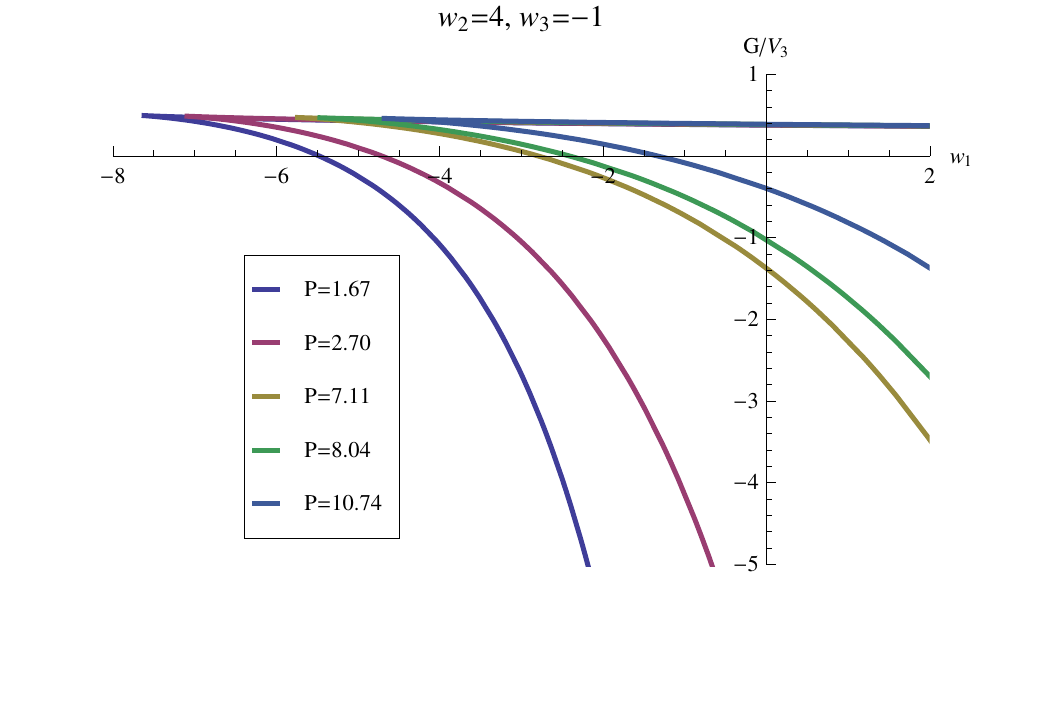}
\caption{The five-dimensional Gibbs free energy as a function of shifted
temperature with different parametrization. The ``swallow tail'' behavior appears when $w_2<0$
and $w_3>0$. }
\end{figure}

Now let us turn to calculate the critical exponents in the massive
gravity black hole system. In the usual thermodynamic system, the
critical exponents $\alpha$, $\beta$, $\gamma$, and $\delta$ are
defined as follows,
\begin{eqnarray}
C_v&\sim&\left(-\frac{T-T_c}{T_c}\right)^{-\alpha},\nonumber\\
\frac{v_g-v_l}{v_c}&\sim&\left(-\frac{T-T_c}{T_c}\right)^{\beta},\nonumber\\
\kappa_T&\sim&\left(-\frac{T-T_c}{T_c}\right)^{-\gamma},\nonumber\\
P-P_c&\sim&\left(v-v_c\right)^{\delta},
\end{eqnarray}
where $C_v$ is the isopyknic heat capacity, $v$ stands for specific
volume, and $\kappa_T$ is isothermal compressibility. The subscript
$c$ stands for the critical point.

For our massive gravity black hole system, the size is determined by
the horizon radius, and we shall use the following expansion parameters
to characterize the critical behavior near the critical point:
\begin{equation}
\tau=\frac{w_1}{w_{1c}}-1,~~~\epsilon=\frac{r_h}{r_{hc}}-1,~~~p'=\frac{P}{P_c}.
\end{equation}
Then we can make the Taylor series expansion near the critical point for
the equation of state (\ref{P}) as
\begin{equation}
p'=1+\frac{8}{3}\tau-\frac{8}{3}\tau\epsilon-\frac{4}{3}\epsilon^3+\mathcal{O}(\tau\epsilon^2,\epsilon^4).
\end{equation}
By using Maxwell's equal area law, we obtain the following equation:
\begin{eqnarray}\label{equalarealaw}
0&=&\int_{\epsilon_l}^{\epsilon_{g}}(\epsilon+1)^3\frac{\mathrm{d}p'}{\mathrm{d}\epsilon}\mathrm{d}\epsilon=\int_{\epsilon_l}^{\epsilon_{g}}(\epsilon+1)^3(-\frac{8}{3}\tau-4\epsilon^2)\mathrm{d}\epsilon\nonumber\\
&=&-\frac{8}{3}\tau(\epsilon_g-\epsilon_l)-\frac{4}{3}(\epsilon_g^3-\epsilon_l^3)-4\tau(\epsilon_g^2-\epsilon_l^2)-\frac{8}{3}\tau(\epsilon_g^3-\epsilon_l^3)+\mathcal{O}(\epsilon^4).
\end{eqnarray}
On the other hand, the pressures of two phases keep the same value
when the phase transition happens:
\begin{equation}\label{pg=pl}
p'|_{\epsilon_l}=p'|_{\epsilon_g}\Rightarrow-\frac{8}{3}\tau(\epsilon_g-\epsilon_l)-\frac{4}{3}(\epsilon_g^3-\epsilon_l^3)+\mathcal{O}(\epsilon^4)=0.
\end{equation}
The above two equations (\ref{equalarealaw}) and (\ref{pg=pl}) have
a unique nontrivial solution $\epsilon_g>0$, $\epsilon_l<0$ when
$\tau<0$. It is easy to find that
\begin{equation}
\epsilon_g-\epsilon_l=2\sqrt{-2\tau+\mathcal{O}(\tau^2)},
\end{equation}
which determines the critical exponent $\beta=1/2$. The isothermal
compressibility can be calculated as follows
\begin{equation}
\kappa_T=-\frac{1}{(\epsilon+1)^3}\left(\frac{\partial(\epsilon+1)^3}{\partial
P}\right)\Big|_{\tau}\propto\left(\frac{\partial
p'}{\partial\epsilon}\right)^{-1}\Big|_{\epsilon=0}=-\frac{3}{8\tau},
\end{equation}
which indicates the critical exponent $\gamma=1$. The difference
between the pressure and its critical value near the critical point
$(p'-1)|_{\tau=0}=-\frac{4}{3}\epsilon^3$ tells us that $\delta=3$.
The isopyknic heat capacity vanishes since the entropy is also
determined by the horizon radius, and then we have $\alpha=0$. The critical
exponents satisfy the following thermodynamic scaling laws,
\begin{eqnarray}
&~&\alpha+2\beta+\gamma=2,~~~\alpha+\beta(1+\delta)=2\nonumber\\
&~&\gamma(1+\delta)=(2-\alpha)(\delta-1),~~~\gamma=\beta(\delta-1),
\end{eqnarray}
which are the same as those in the van der Waals liquid-gas system.

\section{$P$-$V$ criticality of five-dimensional neutral black holes}
In the five-dimensional case, we have $n=3$ and $\mathcal{U}_4=0$. So we can set $c_4=0$ in the metric function. In this section, we will
show that the $c_3m^2$ term in five-dimensional neutral black holes can play a similar role as the charge. We set $Q=0$ for simplicity. The enthalpy (\ref{enthalpy}) becomes
\begin{equation}
H=\frac{3V_3r_h^2}{16\pi}\Big[k+\frac{4\pi
P}{3}r_h^2+\frac{c_0c_1m^2}{3}r_h+c_0^2c_2m^2+\frac{2c_0^3c_3m^2}{r_h}\Big].
\end{equation}
The equation of state is
\begin{equation}
P=\Big(\frac{3}{4}T-\frac{3c_0c_1m^2}{16\pi}\Big)\frac{1}{r_h}-\Big(\frac{3k}{8\pi}+\frac{3c_0^2c_2m^2}{8\pi}\Big)\frac{1}{r_h^2}-\frac{3c_0^3c_3m^2}{8\pi}\frac{1}{r_h^3}.
\end{equation}

\begin{figure}
\includegraphics[width=8cm]{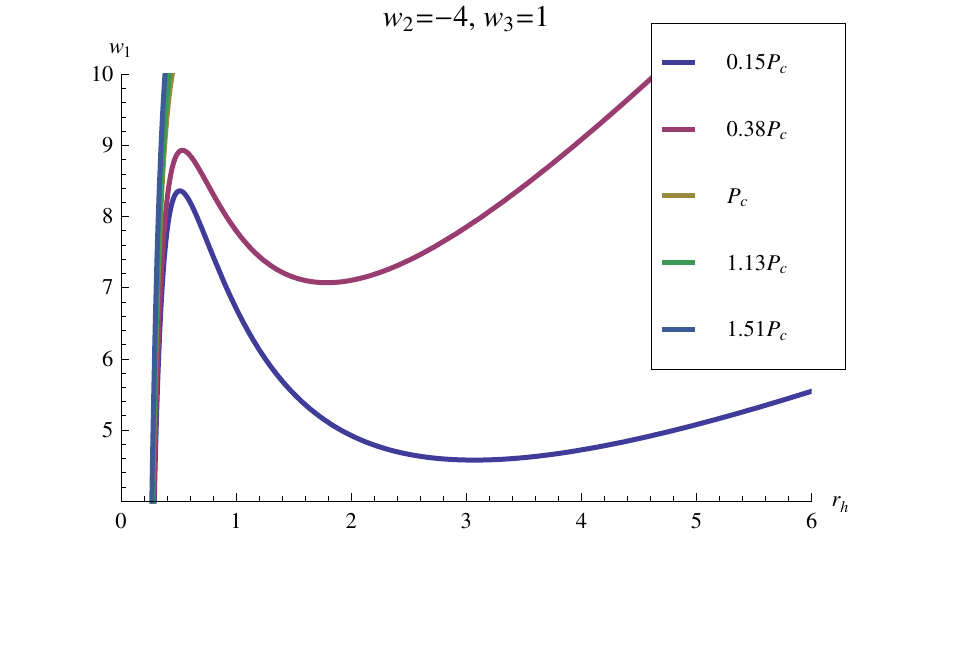}
\includegraphics[width=8cm]{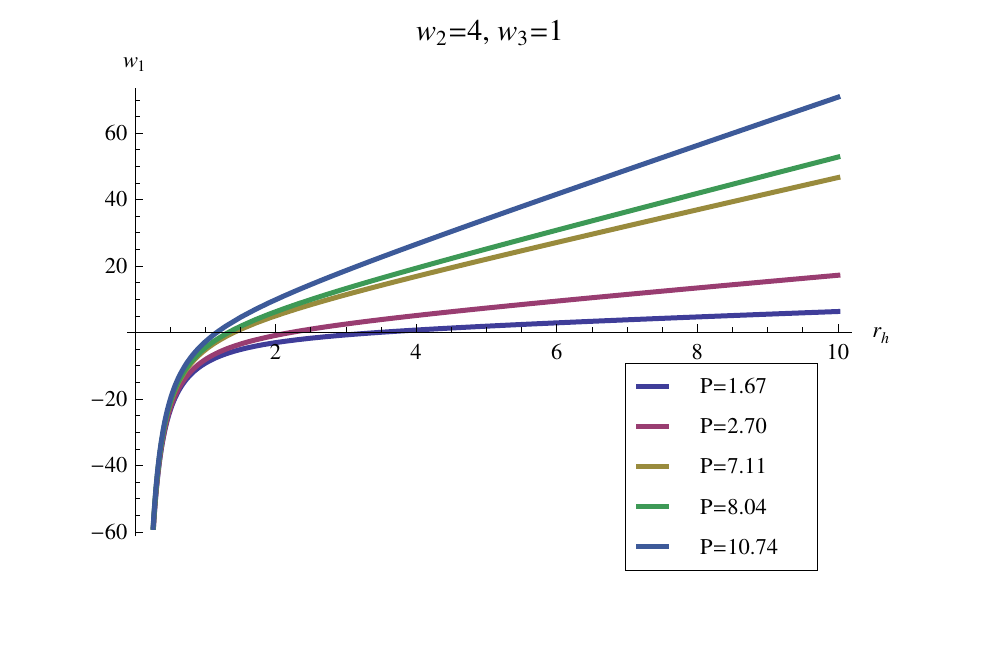}\\
\includegraphics[width=8cm]{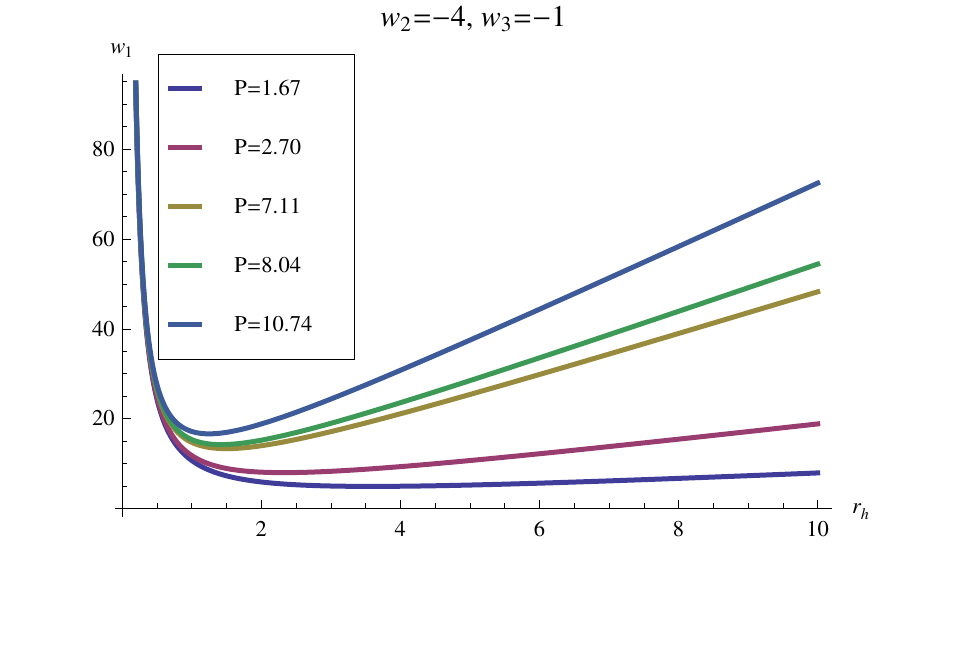}
\includegraphics[width=8cm]{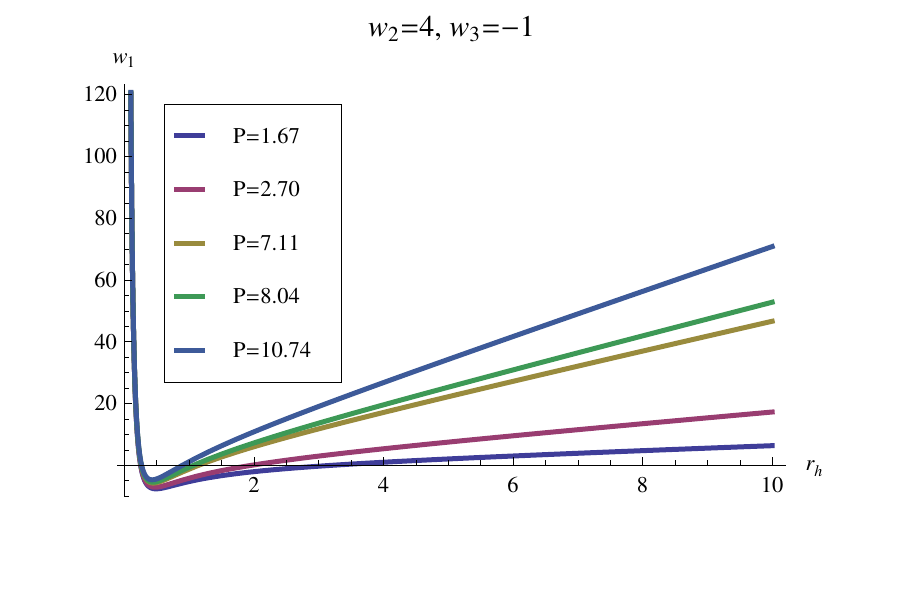}
\caption{The 5-dimensional $w_1$-$r_h$ diagrams with different parametrization. The negative slop indicate the negative heat capacity and hance the thermodynamic instability.}
\end{figure}

As in the four-dimensional case, it is convenient to denote
\begin{equation}\label{w3}
w_3=-\frac{c_0^3c_3m^2}{8\pi},
\end{equation}
so the equation of state can be put in a neat form according to the
definition of $w_1$ and $w_2$ in Eq.(\ref{w124})
\begin{equation}\label{P5}
P=\frac{3w_1/2}{r_h}+\frac{3w_2}{r_h^2}+\frac{3w_3}{r_h^3}.
\end{equation}
The critical point is determined by the vanishing of the first and
second derivative of $P$ with respect to $r_h$. The critical value
of $r_h$, $w_1$, and $P$ can be calculated as follows:
\begin{equation}
r_{hc}=-\frac{3w_3}{w_2},
\end{equation}
\begin{equation}
w_{1c}=\frac{2w_2^2}{3w_3},
\end{equation}
\begin{equation}
P_c=-\frac{w_2^3}{9w_3^2}.
\end{equation}
Similarly, we shall keep $w_2<0$ and $w_3>0$ when the critical behavior
appears, and the critical coefficient in this case is
\begin{equation}
\frac{P_cr_{hc}}{w_{1c}}=\frac{1}{2}.
\end{equation}

By the same token, one can show that the critical exponents in this
five-dimensional case are kept the same as those in four dimensions.

One can draw the phase diagram to see the phase structure near
the critical point. In the Fig. 3 and 4, we give the
$P$-$V(r_h)$ and $G$-$w_1$ diagrams according to the different signs of $w_2$ and $w_3$ in this five-dimensional neutral black hole case. 

The positivity of $w_3$ in equation of state (\ref{P5}) leads to a
divergence of pressure when $r_h\rightarrow0$ which is very similar to the charge effect in four dimensions. And the sign of $w_2$ determines whether there exists a first-order phase transition or not. For a van der Waals like phase transition, $w_2$ must be negatively valued. Interestingly, when $w_3$ is negative, in spite of  the value of $w_2$, there always exists one preiection in the $P$-$V(r_h)$ diagram, and the pressure drops to zero when the horizon radius decreases to a nonzero value. A more reasonable and physical explanation will be explored in the next section.

In this chargeless case, the vacuum solution should also be taken into account. According to the Hamiltonian approach, the vacuum solution must be the one with zero energy (here is the enthalpy), entropy and, hance, the horizon radius. The Gibbs-free energy of the vacuum solution vanishes by the definition (\ref{G}). The first diagram of the Fig. 4 shows a ``swallow tail'' behavior when the pressure is below a critical point. Note that only the ``tail'' may appear in the upper half plane with positive Gibbs-free energy, the thermodynamic stable phase will always be the black hole phase, and the phase space stays the same as the one in four dimensions. This is true in the second diagram as well. In the third and fourth diagram of the Fig. 4 with $w_3=-1$, only the part in the lower half plane indicates a stable black hole solution. The shifted temperature has a minimal value below which no black hole solution exists. In fact, when the temperature drops to a certain value larger than the minimal one, the Gibbs-free energy will become larger than zero and a more stable vacuum will take place. This is a Hawking-Page-like phase transition.    

On the other hand, the thermodynamic instability can also be implied by negatively valued heat capacity. The heat capacity of the balck hole system with fixed pressure and charge can be calculated as
\begin{eqnarray}  
C_{P,Q}&=&\left(\frac{\partial H}{\partial T}\right)_{P,Q}=T\left(\frac{\partial S}{\partial T}\right)_{P,Q}\nonumber\\
&=&T\frac{\mathrm{d} S}{\mathrm{d}r_h}\left(\frac{\partial T}{\partial r_h}\right)^{-1}_{P,Q}.
\end{eqnarray}
We can see that the sign of $C_{P,Q}$ is determined by the sign of $\left(\frac{\partial T}{\partial r_h}\right)^{-1}_{P,Q}$ and, hance, by the sign of $\left(\frac{\partial w_1}{\partial r_h}\right)^{-1}_{P,Q}$. In the Fig. 5, we plot the the $w_1$-$r_h$ diagrams with fixed pressures in this neutral black hole case. We keep the same values of $w_2$ and $w_3$ as in $G$-$w_1$ diagrams. The negative slope of $w_1$-$r_h$ curve indicates a negative heat capacity and, hance, instability. The first diagram of Fig. 5 shows the instability with negative slope when the pressure is below the critical one, which is consistent with the $G$-$w_1$ analysis. The second diagram of Fig. 5 with $w_2=4$ and $w_3=1$ tells us that there is only one stable black hole phase. The last two diagrams of Fig. 5 are also consistent with the $G$-$w_1$ analysis that there exist unstable black hole phases with $C_{P,Q}<0$.

\section{$P$-$V$ criticality of general higher-dimensional black holes}
In this section, we will analyze the $P$-$V$ criticality of charged
black holes in general spacetime dimensions based on the previous
analysis of special cases. According to the Hawking temperature,
i.e., Eq. (\ref{Htemp}), the equation of state for such charged
black holes with general spacetime dimensions is
\begin{eqnarray}\label{nP}
P&=&\Big(\frac{nT}{4}-\frac{nc_0c_1m^2}{16\pi}\Big)\frac{1}{r_h}-\frac{n(n-1)(k+c_0^2c_2m^2)}{16\pi}\frac{1}{r_h^2}-\frac{n(n-1)(n-2)c_0^3c_3m^2}{16\pi}\frac{1}{r_h^3}\nonumber\\
&~&-\frac{n(n-1)(n-2)(n-3)c_0^4c_4m^2}{16\pi}\frac{1}{r_h^4}+\frac{8\pi
Q^2}{V_n^2}\frac{1}{r_h^{2n}}.
\end{eqnarray}
Obviously, we can see that the power of $r_h$ of the $c_i$ terms is
independent of spacetime dimensions, while the charge term is not. This ensure that the contributions of every $c_i$ term in different spacetime dimensions are qualitatively the same. The things that matter for the diversity of the phase structure or $P$-$V$ diagram are the value of $c_im^2$s. For example, the $r_h$ dependence of the
$c_4m^2$ term is $r_h^{-4}$ when $c_4m^2$ is negatively valued and $n>3$; it gives a chargelike contribution as in four spacetime
dimensions. In any spacetime dimension, the charge term plays a
dominant role as $r_h\rightarrow0$ since it has the lowest power of
$r_h$. Note that the charge term has a positive sign, and the
pressure will tend to infinity when $r_h\rightarrow0$ as long as
the black hole is charged.

Physically, an infinite pressure implies the existence of repulsion
interaction. Let us take the van der Waals liquid-gas system as an
example. The equation of state is the van der Waals equation
\begin{equation}\label{vdWeq}
P=\frac{T}{v-b}-\frac{a}{v^2},
\end{equation}
where $v$ is the so-called specific volume, and $a$ and $b$ are positive
constants determined by experiments. In contrast to the ideal gas
with $a=b=0$, constants $a$ and $b$ reflect the attraction and
repulsion force between molecules, respectively. When $b/v$ is not
very large, the van der Waals equation can be written as an Annes
equation with series expansion
\begin{equation}\label{anneseq}
P=\frac{T}{v}+\frac{bT}{v^2}-\frac{a}{v^2}+\frac{2b^2T}{v^3}+\frac{6b^3T}{v^4}+\mathcal{O}(v^{-5}).
\end{equation}
In the above equation (\ref{anneseq}), the pressure is expressed as
a polynomial of $v$, the positive coefficient reflects the repulsion
interactions, and the negative coefficient is related to the attraction
interactions. As to our state equation of charged black holes in
general spacetime dimensions, i.e., Eq. (\ref{nP}), if we identify
the horizon radius $r_h$ as the specific volume, it has a very
similar structure as the Annes equation (\ref{anneseq}). The positive
sign of the charge term implies a repulsion interaction which could be
understood. The value of $c_im^2$s cannot be determined by the
theory itself. This leaves many possibilities which lead to a
plentiful phase structure. For example, if $c_4m^2$ is negatively
valued, then the fourth term in Eq. (\ref{nP}) reflects a repulsion
interaction, and vice versa. Similar properties can be applied to
$c_3m^2$. The sign of the second term of Eq. (\ref{nP}) is determined by
$k+c_0^2c_2m^2$, which is a joint effect of the horizon topology and
$c_2m^2$. In other words, at least in the thermodynamic sense, the
appearance of $c_2m^2$ mist the contribution of horizon topology.
This can also been seen from the metric function (\ref{f}) where the
constant term is $k+c_0^2c_2m^2$. Thus, unlike with the charged AdS
black hole in GR, a hyperbolic or spherical horizon topology does
not necessarily imply a repulsion or attraction interaction. It
depends on the sign of $k+c_0^2c_2m^2$. The appearance of $c_1m^2$
seems exotic since no such term appears in the usual Annes equation
(\ref{anneseq}). At present, we can only understand it as a
correction to the Hawking temperature.

Now we can conclude that the appearance of the critical behaviors or
the first-order phase transitions is the results of competition of
repulsion and attraction interactions between some unknown degree of
freedom. For such phase transitions, an infinite repulsion
interaction is necessary when the horizon radius tends to zero.

\section{Conclusions and Discussion}
In this paper, we have found that in the context of massive gravity,
there also exists the van der Waals-like phase transition in the
extended phase space of charged AdS black holes when the
cosmological constant is identified as the thermodynamical pressure.
For such identification, the black hole mass must be viewed as
thermal enthalpy rather than the internal energy of the gravitational
system. Then we use the standard thermodynamic identities to obtain
the entropy, volume, and electric potential. The entropy keeps the
one quarter of the horizon area law as in GR. In contrast with black
holes in GR, the phase structure becomes richer due to the
appearance of the massive potential associated with the graviton mass.
There are four terms in the massive potential. As to the spatial
reference metric black holes, the contribution of these four terms
in some thermodynamic quantities like enthalpy or temperature is
dimensional dependent. The Smarr relation is modified by this
massive potential.

In the four-dimensional case, only the first and second terms appear
in the thermal enthalpy as well as the equation of state. In the
canonical ensemble, the van der Waals-like phase transition happens
when $k+c_0^2c_2m^2>0$. In this first-order phase transition, we
calculate the critical exponents and find that these are the same
as in the van der Waals liquid-gas phase transition. It is worth
mentioning that we have used the shifted temperature rather than the real
one to characterize the critical point. In this sense, we find the same
critical coefficient as lin the iquid-gas phase transition. The figure of
Gibbs-free energy as a function of shifted temperature for constant
pressure shows the ``swallow tail'' behavior when $k+c_0^2c_2m^2>0$.
The ``tail'' characterizes the unstable state with larger values of
Gibbs-free energy for constant temperature, pressure, and electric
charge.

In the five-dimensional case, the third term in the massive
potential contributes. In order to see the effect of the term
$c_3m^2$, we simply set $Q=0$. There also can undergo a van der
Waals-like phase transition if, and only if $k+c_0^2c_2m^2>0$ and
$c_3m^2<0$. Qualitatively, there are four possible shapes in the
$P$-$V$ diagram according to the different sign of $k+c_0^2c_2m^2$ and
$c_3m^2$. When $c_3m^2>0$, there always exists a Hawking-Page-like phase transition at a certain temperature from the black hole state to the vacuum state.

Generally, we can express the thermodynamical pressure as polynomial
in terms of horizon radius. By contrasting with the Annes equation, we
find that it is natural to identify the horizon radius as the
specific volume. As a polynomial, the positive coefficient reflects
the repulsion interactions and the negative coefficient is related to
the attraction interactions. The appearance of the critical
behaviors or the first-order phase transitions are the result of the
competition of repulsion and attraction interactions between some
unknown degree of freedom. In particular, the $c_2m^2$ term will mix
the contribution of horizon topology and the $c_1m^2$ term can be
viewed as a correction to the Hawking temperature.

Finally, we would like to mention that the black hole solution in
massive gravity is tightly related to the choice of the reference
metric. Whether the thermodynamic properties depend on the choice of
the reference metric is another interesting topic that deserves future study.

\section{Acknowledgments}
We would like to thank Professor Rong-Gen Cai for his valuable disscussion and comments. This work was supported in part by the National Natural Science Foundation of China under Grants No. 11205148, No. 11235010, and No. 11105004, the Fundamental Research Funds for the Central Universities under Grant No. NS2015073, and Shanghai Key Laboratory of Particle Physics and Cosmology under Grant No. 11DZ2260700.


\end{document}